\documentclass[12pt,twoside]{article}
\usepackage{fleqn,espcrc1,epsf}

\usepackage{graphicx}

\newcommand{\Gf}{GeV/fm$^3$ }
\newcommand{\Tc}{T_{\rm c}}
\newcommand{\vp}{\langle v_\perp\rangle}
\newcommand{\mt}{m_{\perp}}

\title{The Little Bang: Searching for quark-gluon matter in 
       relativistic heavy-ion collisions}

\author{Ulrich Heinz\thanks{On leave from Institut f\"ur Theoretische Physik, 
                Universit\"at Regensburg, D-93040 Regensburg, Germany.
                Work supported in part by DFG, GSI, and BMBF. Email address:
                {\tt ulrich.heinz@cern.ch}}
        \\[1ex]
        Theoretical Physics Division, CERN,
        CH-1211 Geneva 23, Switzerland}%
       
\begin{document}

\maketitle

\begin{abstract}
I review the status of the search for quark-gluon plasma in 
relativistic heavy-ion collisions. The available data provide strong
evidence for the "three pillars of the Little Bang model": strong 
radial expansion of the collision fireball with Hubble-like scaling, 
thermal hadron emission and primordial hadrosynthesis. It is argued 
that the initial state of the reaction zone exhibits features which 
cannot be understood with conventional hadronic dynamics, but are 
consistent with the formation of deconfined quark-gluon matter.
\end{abstract}

\section{PROLOGUE}
On Feb. 10, 2000, CERN announced officially \cite{CERN} that (I 
paraphrase) ``compelling evidence now exists for the formation of a 
new state of matter at energy densities about 20 times larger than 
that in the center of atomic nuclei and temperatures about 
100000 times higher than in the center of the sun. This state exhibits
characteristic properties which cannot be understood with conventional
hadronic dynamics but which are qualitatively consistent with 
expectations from the formation of a state of matter in which quarks 
and gluons no longer feel the constraints of color confinement.''
I will here explain the scientific arguments \cite{HJ00} which led 
to this conclusion. The announcement was not triggered by a single 
experimental discovery which had just happened, but emerged from a 
painstaking analysis of many different sets of data collected during 15 
years of heavy-ion collision experiments at the CERN SPS, especially of
results obtained during the last five years with the 158\,$A$\,GeV/$c$ 
$^{207}$Pb beam and reported over the last 24 months \cite{conferences}. 
It happened at a time when the Relativistic Heavy Ion Collider RHIC 
at BNL was about to turn on and CERN was taking stock of what had 
been achieved so far with the up to then highest energy heavy-ion 
beams available. The announcement gave credit to the achievements of 
an international community of almost 500 enthusiastic physicists from 
all over the world who had initiated and driven to success the 
heavy-ion program at CERN, participating in seven large and several 
smaller experiments with complementary goals in order to search for 
the creation of the theoretically predicted ``quark-gluon plasma'' 
(QGP) in heavy-ion collisions. 

The reaction of the international press to this announcement was very 
strong but the reports were not always accurate. Several stated that 
CERN had claimed the ``discovery of quark-gluon plasma''. This was not 
true: The CERN press release \cite{CERN} and the scientific document 
on which it was based \cite{HJ00} had been formulated very carefully 
and made a clear and conscious distinction between ``evidence for a new 
state of matter'' (which we claimed) and ``discovery of QGP'' (which we 
didn't). I will also report on the remaining open questions, trying to 
point out (as I did in my talk on Feb. 10 \cite{mytalk}) what is needed 
to turn the present {\em evidence} into incontrovertible {\em proof}. 
Several crucial issues can only be resolved by heavy-ion experiments at 
the higher energies provided by RHIC and LHC.

\section{THE QUARK-HADRON TRANSITION}

The QCD phase diagram \cite{Raja} features a transition from a gas of
hadronic resonances (HG) at low energy densities to a quark-gluon 
plasma (QGP) at high energy densities. The critical energy density 
$\epsilon_{\rm c}$ is of the order of 1\,\Gf. It can be reached by 
either heating matter at zero net baryon density to a temperature of 
about $\Tc\approx 170$\,MeV, or by compressing cold nuclear matter
to baryon densities of about $\rho_{\rm c}\sim 3-10\,\rho_0$ (where 
$\rho_0=0.15$\,fm$^{-3}$ is the equilibrium density), or by combinations 
thereof. A simple version of the phase diagram is shown in Figure\,1.
 
\begin{minipage}[c]{10cm}
\vspace*{0.5cm}
\hspace*{-0.5cm}
  \epsfxsize 9cm
  \epsfbox{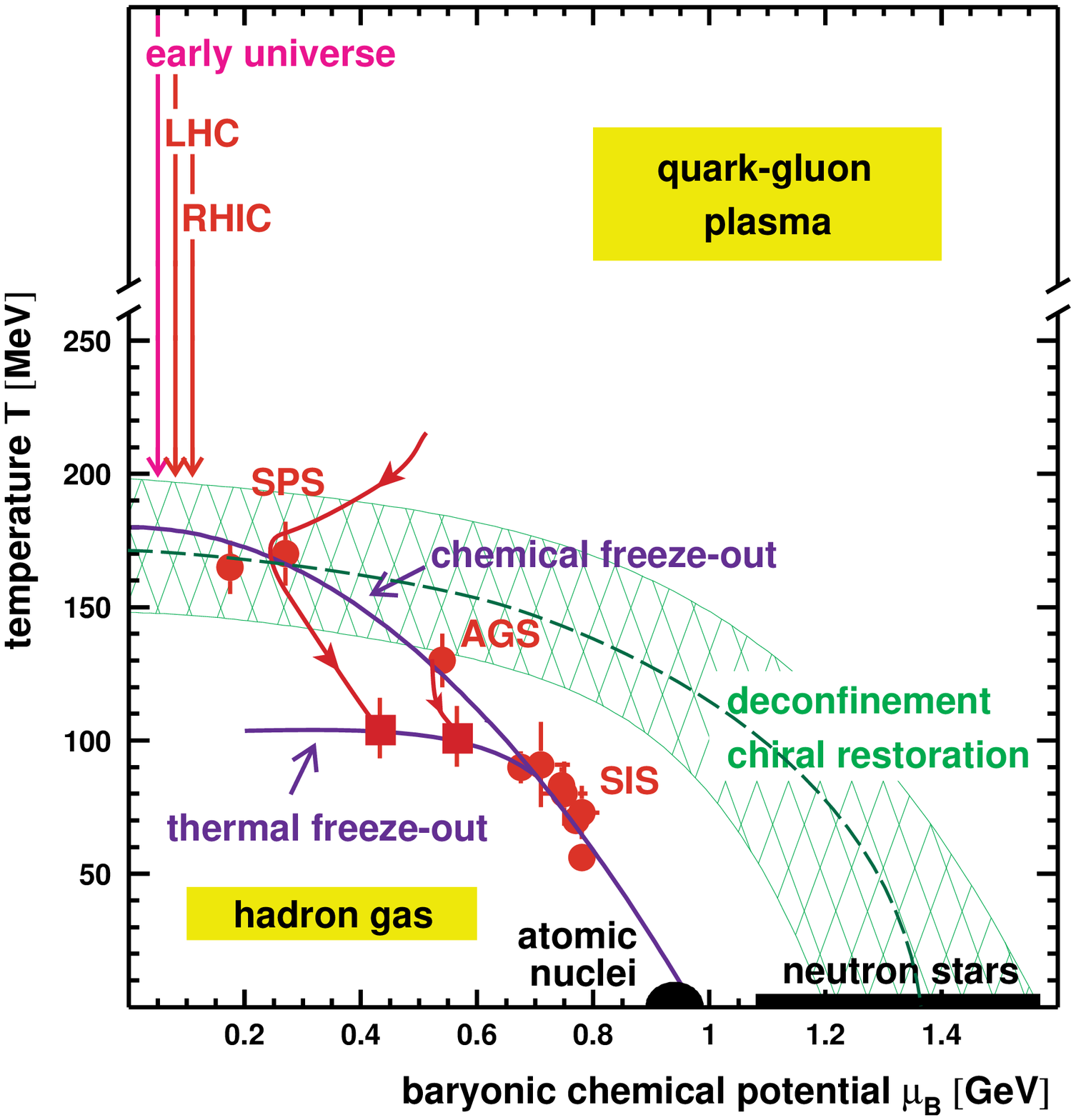}
\end{minipage}
\hfill\hspace*{-1.5cm}
\parbox[c]{6.5cm}{\vspace*{0cm}
    Figure~1. Sketch of the QCD phase diagram, temperature $T$ vs. the 
    baryon chemical potential $\mu_{\rm B}$ associated with net baryon 
    density $\rho_{\rm B}$. The cross-hatched region indicates the 
    expected phase transition and its present theoretical uncertainty,
    with the dashed line representing its most likely location. Lines
    with arrows indicate expansion trajectories of thermalized matter 
    created in different environments. (The lines labelled ``RHIC'', 
    ``LHC'' and ``early universe'' should be much closer to the 
    temperature axis but would then have been difficult to draw.) For 
    a discussion of the chemical and thermal freeze-out lines and the 
    location of the data points see text.
\label{F1}}
\vspace*{0.2cm}

By colliding heavy ions at high energies, one hopes to heat and compress 
hadronic matter to energy densities above $\epsilon_{\rm c}$. At lower 
beam energies (SIS @ 1\,$A$\,GeV/$c$), the nuclei are stopped and the 
nuclear matter is compressed and moderately heated. At higher beam 
energies one reaches higher temperatures, but since the colliding nuclei 
are no longer completely stopped, the baryon chemical potential of the 
matter created at rest in the center-of-momentum system decreases (AGS 
@ 10\,$A$\,GeV/$c$ and SPS @ 160\,$A$\,GeV/$c$). At the heavy ion 
colliders RHIC ($\sqrt{s}=200\,A$\,GeV) and LHC ($\sqrt{s}=5500\,A$\,GeV) 
the baryon chemical potential of the reaction fireball is so small that
one essentially simulates baryon-free hadronic matter, very much like the
expanding early universe. If the matter thermalizes quickly at energy
densities above $\epsilon_{\rm c}$, it will pass through the quark-hadron
phase transition as the collision fireball expands and cools.

Along the temperature axis at $\mu_{\rm B}=0$ our knowledge of the 
QCD phase diagram is based on hard theory (lattice QCD), but for
nonzero baryon density we must rely on models interpolating between 
low-density hadronic matter, described by low-energy effective theories, 
and high-density quark-gluon plasma, described by perturbative QCD. 
The theoretical uncertainties at high-baryon densities are 
thus difficult to quantify and relatively large (typically 
${\cal O}(30-50\%)$). At zero baryon density, the situation is 
much cleaner: numerical simulations of QCD with 3 dynamical light quark 
flavors on the lattice are now available, and the systematic errors due 
to lattice discretization and continuum extrapolation are controllable 
and beginning to get small \cite{Karsch}. The critical temperature 
$\Tc$ for real-life QCD is predicted as $T_{\rm c}
 \approx 170$\,MeV\,$\pm15\%$ \cite{Karsch,priv}. Near $T_{\rm c}$ the 
energy density in units of $T^4$ changes dramatically by more than a 
factor of 10 within a very narrow temperature interval (see Figure 2). 
Above $T\simeq 1.2\, \Tc$, $\epsilon/T^4$ appears to settle at about 
80\% of the Stefan-Boltzmann value for an ideal gas of non-interacting 
quarks and gluons. 

\begin{minipage}[c]{10cm}
\vspace*{-0.2cm}
\hspace*{-1.3cm}
  \epsfxsize 10.5cm
  \epsfbox{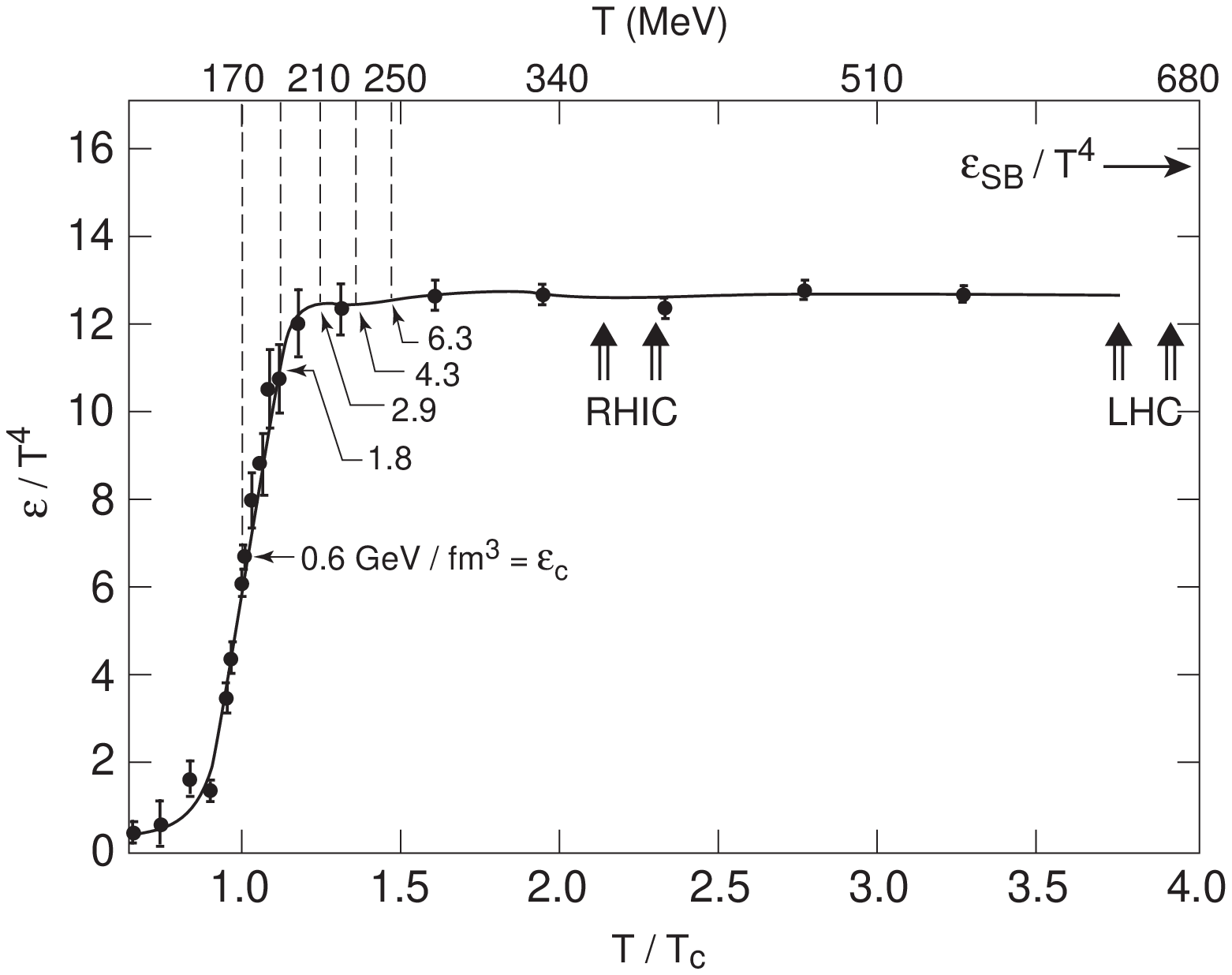}
\end{minipage}
\hfill\hspace*{-1.5cm}
\parbox[c]{6.5cm}{\vspace*{-0.4cm}
    Figure~2. Energy density in units of $T^4$ for QCD with 3 dynamical
    light quark flavors of mass $m/T = 0.4$ \cite{priv}. For two light 
    and a heavier strange quark flavor ($m/T=1$) the energy density 
    decreases by $\leq 15\%$ \cite{priv}. The lower horizontal axis 
    gives $T/\Tc$, the upper one converts $T$ to MeV using 
    $\Tc=170$\,MeV. For various values of $T$ absolute values for the 
    energy density $\epsilon$ in \Gf are indicated, using the same $\Tc$. 
\label{F2}}
\vspace*{-0.3cm}

According to Figure\,2 only about 600 MeV/fm$^3$ of energy density 
are needed to make the transition to deconfined quark-gluon matter. 
It is, however, very expensive to reach temperatures well above $\Tc$:
an initial temperature of 220 MeV, about 30\% above $\Tc$,
already requires an initial energy density $\epsilon\simeq 3.5$\,\Gf, 
about 6 times the critical value. We will see that this severely
limits the reach of the CERN SPS into the new QGP phase: only the 
region at and slightly above $\Tc$ can be probed. We are ``living 
at the edge'' (W.\,Zajc)! To really penetrate deep into the new 
phase requires the much higher energy densities which become 
accessible with the heavy-ion colliders RHIC and LHC. 

At $\Tc$ two phenomena happen simultaneously \cite{KL94}: color
confinement is broken, i.e. colored degrees of freedom can propagate 
over distances much larger than the size of a hadron, and the 
approximate chiral symmetry of QCD, which is spontaneously broken at
low temperatures and densities, gets restored. Both effects are 
important since they significantly accelerate particle production: 
the liberation of gluons in large densities opens up new gluonic 
production channels, and the threshold for quark-antiquark pair 
production is lowered since the quarks shed a large fraction of 
their constituent mass which is dynamically generated by their 
interaction with the quark condensate characterizing the 
spontaneously broken chiral symmetry at low densities. 

Much less important for the dynamics of heavy-ion collisions is the 
actual {\em order of the phase transition}: depending on the value of 
the strange quark mass, existing lattice calculations allow for anything 
between a smooth crossover, similar to the one seen in Figure 2, and a
weakly first order transition with a latent heat which is small on the
vertical scale of Figure 2. The resulting differences in the dynamical
evolution of the expanding fireball are almost certainly unmeasurable,
except if the phase diagram features a critical point and the experiments
pass close to it \cite{Raja}.

\section{RECONSTRUCTING THE LITTLE BANG}

As the two nuclei hit each other with full speed, a superposition
of nucleon-nucleon (NN) collisions occurs. What is different 
compared to individual NN collisions is that (i) each nucleon may 
scatter several times, and (ii) the liberated partons from 
different NN collisions rescatter with each other even before 
hadronization, as do the secondary hadrons produced in different 
NN collision. Both features change the particle production 
{\em per participating nucleon}. But only the rescattering processes
can lead to a state of local thermal equilibrium, by redistributing the
energy lost by the beams into the statistically most probable 
configuration. These rescatterings result in thermodynamic pressure 
acting against the outside vacuum which causes the reaction zone to 
expand collectively. The expansion cools and dilutes the fireball below 
the critical energy density of the quark-hadron transition, at which 
point hadrons are formed from the quarks and gluons (hadronization).
Further interactions between these hadrons cease once their average 
distance exceeds the range of the strong interactions: the hadrons
``freeze out''.

The strong interactions among the partons and hadrons before freeze-out 
wipe out most information about their original production processes.
The extraction of information about the interesting hot and dense 
early collision stage thus requires to exploit features which are
either established early and not changed by the rescattering and 
collective expansion or can be reliably back-extrapolated. 
Correspondingly one classifies the observables into two classes, 
{\em early} and {\em late} signatures. (A comprehensive review of
QGP signatures with a complete list of references can be found in
\cite{Bass99}.)

The conceptually cleanest early signatures are the directly produced 
real and virtual photons (i.e. not those resulting from hadron decays 
after freeze-out) since photons show no strong interaction and directly 
escape from the fireball (virtual photons materialize as $e^+e^-$ or 
$\mu^+\mu^-$ pairs). They are emitted throughout the expansion, but
their production is expected to be strongly weighted towards the
hot and dense initial stages. Unfortunately, direct photons are 
rare, and the experimental background from hadronic decay photons 
after freeze-out is enormous. 

Another early signature are hadrons made of charmed quarks. At SPS 
energies, $c\bar c$ pairs can only be created in the primary NN 
collisions, because the secondary scatterings are not energetic 
enough to overcome the $c\bar c$ threshold. The only thing that 
can happen to them is a redistribution in phase-space, changing the 
relative amounts of mesons with hidden ($c\bar c$ charmonium) and open 
charm ($\bar c q$ and $\bar q c$). It was shown that such a 
redistribution is much easier in the color-deconfined QGP phase than 
by reinteraction of charmed particles with other hadrons; in this way 
charm-redistribution becomes also an early signature. But also charm 
production is a very rare process at SPS energies and below.   

Hadrons made of up, down and strange quarks, which can be relatively
easily produced and destroyed in all stages of the fireball expansion 
(I'll slightly qualify this statement further below with regard to 
strange hadrons), are {\em late signatures}. They provide only 
{\em indirect} information about the early collision stages. But they 
are very abundant, forming more than 99.99\% of all emitted particles, 
and can thus be measured very accurately. We'll thus use these 
{\em late} signals to reconstruct the Little Bang and then check
the consistency of the resulting picture with the less detailed 
direct information from the {\em early} signals.  

I would like to note that this procedure is quite analogous to the
reconstruction of the cosmological Big Bang from observations. The three
observational pillars of the Big Bang theory are the Hubble expansion  
(which goes on until today), the cosmic microwave background of
thermal photons (which decoupled when our universe was about 300,000 
years old), and the measured abundances of light atomic nuclei (which 
reflect the primordial nucleosynthesis until about 3 minutes after the 
Big Bang). Almost everything before that time is based on a theoretical
back-extrapolation using Einstein's equations and the Standard Model of
particle physics and not on direct astrophysical observations. The only 
exception is the observed baryon number asymmetry which was presumably
established very early but which is still not entirely understood. 

\section{INITIAL CONDITIONS}

Before starting to analyze the final freeze-out stage of the collision
using the {\em late signatures}, it is useful to get an idea about 
the global conditions of the fireball formed shortly after impact.
Following Bjorken \cite{Bj} it is possible to estimate the initially 
{\em produced} energy density by measuring the total transverse energy 
$E_T$ (excluding the fraction due to motion along the beam 
direction) and putting it into an estimated initial volume of the 
reaction zone. Assuming boost-invariant longitudinal expansion (which 
is expected to hold at high energies \cite{Bj} and for which 
evidence exists in the midrapidity region even at SPS energies
\cite{NA49HBT}) we can identify the space-time rapidity $\eta = 0.5
\ln[(t{+}z)/(t{-}z)]$ with the momentum-space rapidity 
$y=0.5\ln[(E{+}p_z)/(E{-}p_z)]$ and write \cite{Bj}
 \begin{equation}
 \label{bj}
  \epsilon_{\rm Bj}(\tau_0) =
  {1\over \pi R_{\rm rms}^2}\, {1\over 2\tau_0} \, {dE_T\over dy}\, .
 \end{equation}
Inserting $\pi R_{\rm rms}^2=63$\,fm$^2$ for the overlap area 
of two Pb nuclei colliding at zero impact parameter, choosing 
$\tau_0=1$\,fm/$c$ to evaluate the length $2\tau dy=2\tau d\eta$ of a 
slice of a cylinder of width $d\eta$ at midrapidity $y=\eta=0$, and using
$dE_T/dy(y=0)\approx 400$\,GeV for central Pb+Pb collisions \cite{epsilon0}
(``very'' central collisions even give up to 10\% more) one obtains
 \begin{equation}
 \label{eps}
  \epsilon_{\rm Bj}^{\rm Pb+Pb}(1\,{\rm fm}/c) =
  3.2 \pm 0.3\, {\rm GeV/fm}^3\,.
 \end{equation}

As QGP searchers we thus play in the right ball-park; if the matter
were already thermalized after 1\,fm/$c$ (which at this point we don't 
know yet), Figure\,2 tells us that the initial temperature would have 
been $T_0\simeq 210-220$\,MeV. Similar initial energy densities
are obtained from a detailed phase-space analysis of the hadronic 
freeze-out state after back-extrapolation to the time before the onset
of transverse expansion \cite{H97}. Perhaps more importantly, initial 
energy densities between 2 and 10\,\Gf during the first 1-2 fm/$c$ are
obtained \cite{URQMD} even in approaches which avoid quarks and gluons 
as relevant degrees of freedom, like the URQMD code which uses hadronic 
strings and resonances to describe particle production and rescattering 
and is tuned to NN data. Indeed, for Pb+Pb collisions at the SPS, 
URQMD predicts that the {\em prehadronic component} (strings) dominates 
over the produced hadrons for nearly 8 fm/$c$ \cite{URQMD}! This further 
strengthens our expectation to see new physics at the SPS.  
 
\section{THERMAL FREEZE-OUT: AN EXPLODING THERMAL FIREBALL}

The measured hadron spectra contain two pieces of information: (i) Their
normalization, i.e. the {\em yields and abundance ratios}, provides 
the chemical composition of the fireball at the ``chemical freeze-out'' 
point (i.e. when the hadron abundances freeze out); this yields 
information in particular about the degree of chemical equilibration, to
be discussed in Section 7. (ii) The hadronic {\em momentum spectra} provide 
information about thermalization of the momentum distributions and 
collective flow. The latter is caused by thermodynamic pressure 
(resulting from intense rescattering among the consituents) and thus 
reflects, in a time-integrated way, the equation of state of the 
fireball matter. The expansion rate at ``thermal freeze-out'' (i.e. 
along the last-scattering hypersurface which marks the decoupling of 
the momenta) provides the Little Bang analogue of the Hubble constant 
for the Big Bang, while the corresponding freeze-out temperature 
parallels the temperature of the cosmic microwave background at 
the point of photon decoupling.

As in cosmology, where the photon temperature is affected by the 
Hubble expansion and {\em red-shifted} from originally 3000 K 
to 2.7 K today, the hadronic momentum spectra are affected by the 
collective expansion of the collision fireball. Only in this case the 
expansion occurs in three dimensions, and the fireball is observed from 
outside, resulting in a {\em blue-shift} of the apparent temperature 
of the spectra. Since the longitudinal expansion is ambiguous in that 
it is difficult to assess which fraction of the finally observed 
longitudinal flow is generated by hydrodynamic pressure and how much 
is a result of incomplete stopping of the two nuclei, we concentrate 
on {\em transverse} flow, reflected in the transverse mass 
($\mt\,{=}\,\sqrt{m^2{+}p_\perp^2}$) spectra. This type of flow is  
only created {\em after} impact.

Roughly, if rescattering among the fireball constituents results 
in thermalization and collective flow, the shapes of {\em all} 
hadronic $\mt$-spectra can be characterized by just two numbers: 
the temperature $T_{\rm f}$ and the mean transverse flow 
velocity $\vp$ at freeze-out. More exactly, this is only true if all 
hadrons decouple simultaneously (i.e. their rescattering cross 
sections are similar), and the form in which the spectra are 
characterized by $\vp$ may depend on the flow velocity and density 
profiles. For this presentation I will neglect the latter 
subtlety; the former can be checked experimentally.

In the relativistic region, i.e. for $\mt > 2 m_0$, the rest masses 
can be neglected, and the effect of flow on the spectra is given by 
the simple blue-shift formula \cite{LHS90}
 \begin{equation}
 \label{blue}
   T_{\rm slope} = T_{\rm f} \sqrt{1+\vp\over 1-\vp}\,.
 \end{equation}
At large $\mt$ {\rm all} hadron spectra should have the same 
inverse slope $T_{\rm slope}$ (this tests ther\-ma\-lization) but
measuring it does not allow to separate thermal from collective motion. 

In the nonrelativistic region $\mt < 2 m_0$, on the other, flow does
couple to the rest mass: for a linear transverse flow velocity 
profile and a Gaussian transverse density profile one finds exactly
\cite{CL95,coal}
 \begin{equation}
 \label{flow}
   T_{\rm slope} = T_{\rm f}+ {\textstyle{1\over 2}} m_0 \vp^2\,.
 \end{equation}

Such an approximately linear rest mass dependence is indeed observed.
Figure 3 shows clearly that the spectra contain a collective flow 
component; inverse slopes of 300 MeV or more as seen e.g. for the 
protons can obviously not be interpreted as hadronic temperatures 
(see Figures 1 and 2). There is some scatter between the data from 
different experiments, partly due to different kinematic regions for 
fitting the slope. The pion spectra are particularly troublesome due 
to the strong deformation at low $\mt$ from resonance decay 
contributions and Coulomb effects. Strictly speaking, the pions should 
not be included in this plot since they are never non-relativistic in 
the accessible region.

\begin{minipage}[c]{10cm}
\vspace*{0.3cm}
\hspace*{-0.5cm}
  \epsfxsize 9cm \epsfysize 6.5cm
  \epsfbox{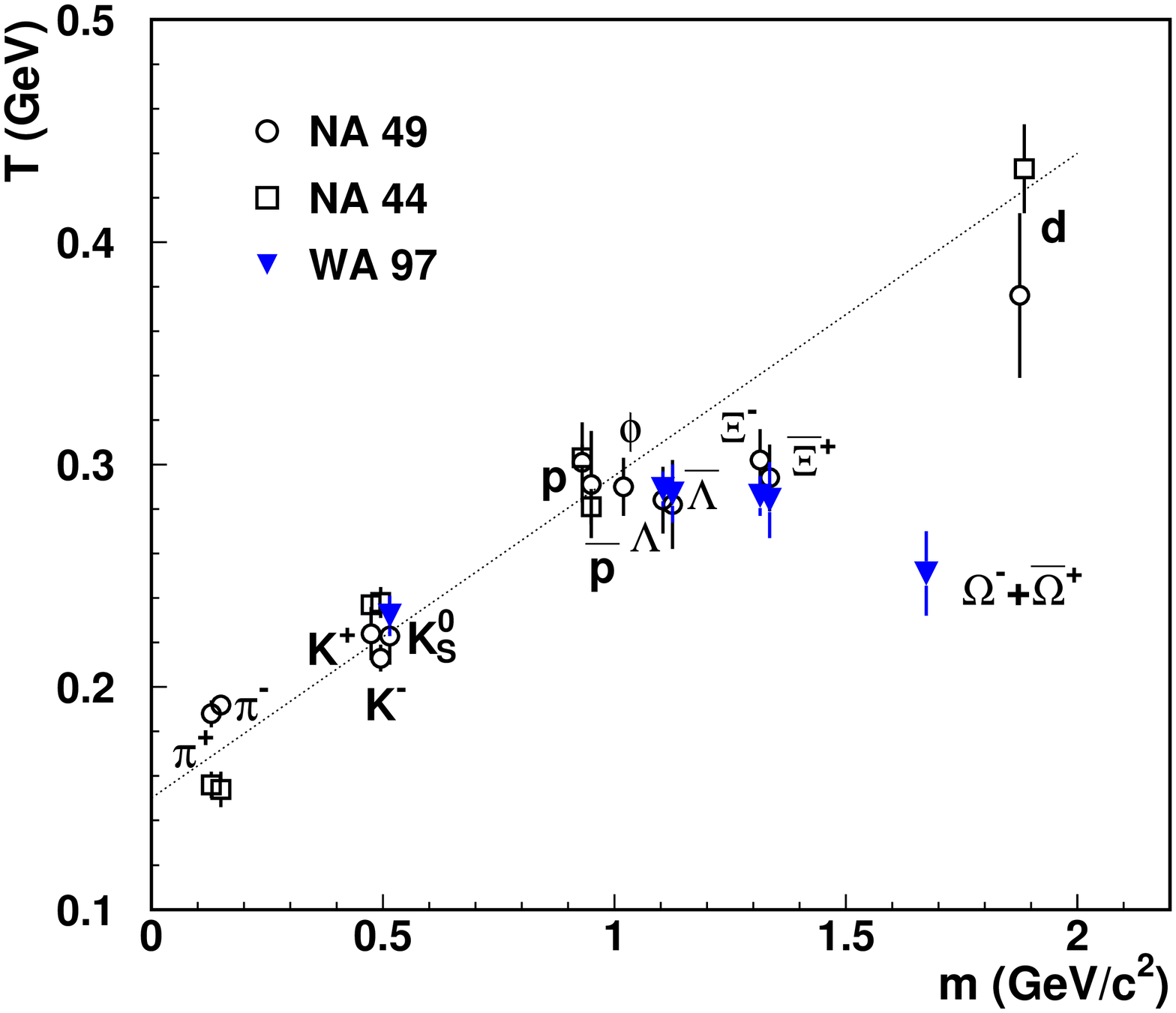}
\end{minipage}
\hfill\hspace*{-1.5cm}
\parbox[c]{6.5cm}{\vspace*{0cm}
    Figure~3. Inverse slopes $T\equiv T_{\rm slope}$ of the measured 
    $\mt$-distributions from 158 $A$\,GeV/$c$ Pb+Pb collisions for 
    various hadron species, plotted against their rest mass 
    \cite{WA97spec}. Results from different experiments are labelled 
    by different symbols. The line is described in \cite{WA97spec}.
\label{F3}}
\vspace*{0.2cm}

The deuterons don't really belong here either: they are too fragile 
to be considered part of the equilibrium ensemble. In the thermal state 
they are continually broken up by collisions, re-forming only at 
freeze-out by coalescence of protons and neutrons. That they fit into
the systematics of Figure 3 is rather a test of the coalescence mechanism
by which the deuterons inherit the temperature and flow from 
their parent nucleons \cite{coal}. In fact, the way they fit 
makes an important statement about the transverse profile of the 
fireball at freeze-out: for a Gaussian density distribution the deuteron
slope should be identical to that of the parent nucleons, whereas the 
observed larger inverse deuteron slope requires a more box-like density 
distribution, with more weight at larger flow velocities \cite{coal,NBI}.

An interesting exception to the trend is provided by the $\Omega$: it's 
spectrum is considerably steeper. This reflects their earlier kinetic
freeze-out, due to an absence of strong scattering resonances with the 
dominating pion fluid \cite{Hecke} which are essential for the kinetic 
re-equilibration of the other hadron species.

After all these {\em caveats} it is clear that a very accurate 
determination of $\vp$ from the slope in Figure\,3 alone is not possible.
But transverse flow also affects other observables, in particular
two-particle correlations in momentum space, like the quantum 
statistical Bose-Einstein (Hanbury Brown-Twiss) correlations between 
identical bosons, or the correlations due to ``soft'' final state 
interactions among the particles after their last ``hard''
scattering which e.g. cause the coalescence of two nucleons into a 
deuteron. In both cases correlations occur only between particles 
which are close in phase-space; by measuring the width of the 
correlation in momentum space one can thus estimate the size
of the emitting source in coordinate space. Collective expansion tends
to reduce the size of the regions within which particles can develop
such correlations; thermal motion, controlled by the thermal velocity 
$\sim\sqrt{T/M_\perp}$, smears out the flow velocity 
gradients and thus acts in the opposite direction. This leads to 
a characteristic dependence of the size of the effective emission 
region for correlated pairs on their transverse mass $M_\perp$; 
its transverse size is controlled by the transverse flow velocity,
$\vp$ as shown by the approximate formula 
\cite{HBT,CL95,coal}  
 \begin{equation}
 \label{Rp}
   R_\perp^2 \approx {R^2 \over 1+ \xi \vp^2 (M_\perp/T_{\rm f})}\,.
 \end{equation}
Here $\xi={\cal O}(1)$ is a model-dependent factor which depends on
the flow velocity and density profiles, and $R$ characterizes the 
geometric transverse size of the fireball at freeze-out.

The formula (\ref{Rp}) predicts that the effective radius $R_\perp$ 
extracted from the correlations scales only with the transverse mass 
$M_\perp$ of the particle pair, irrespective of the rest masses of the
individual particles. This is borne out by experiment, see Figure\,4.
Similar results, consistent with those shown in Figure~4, were obtained
by the NA49 \cite{NA49HBT} and NA52 collaborations \cite{NA52}.

\begin{minipage}[c]{10cm}
\vspace*{-0.2cm}
\hspace*{0.6cm}
  \epsfxsize 7.3cm \epsfysize 9.5cm
  \epsfbox{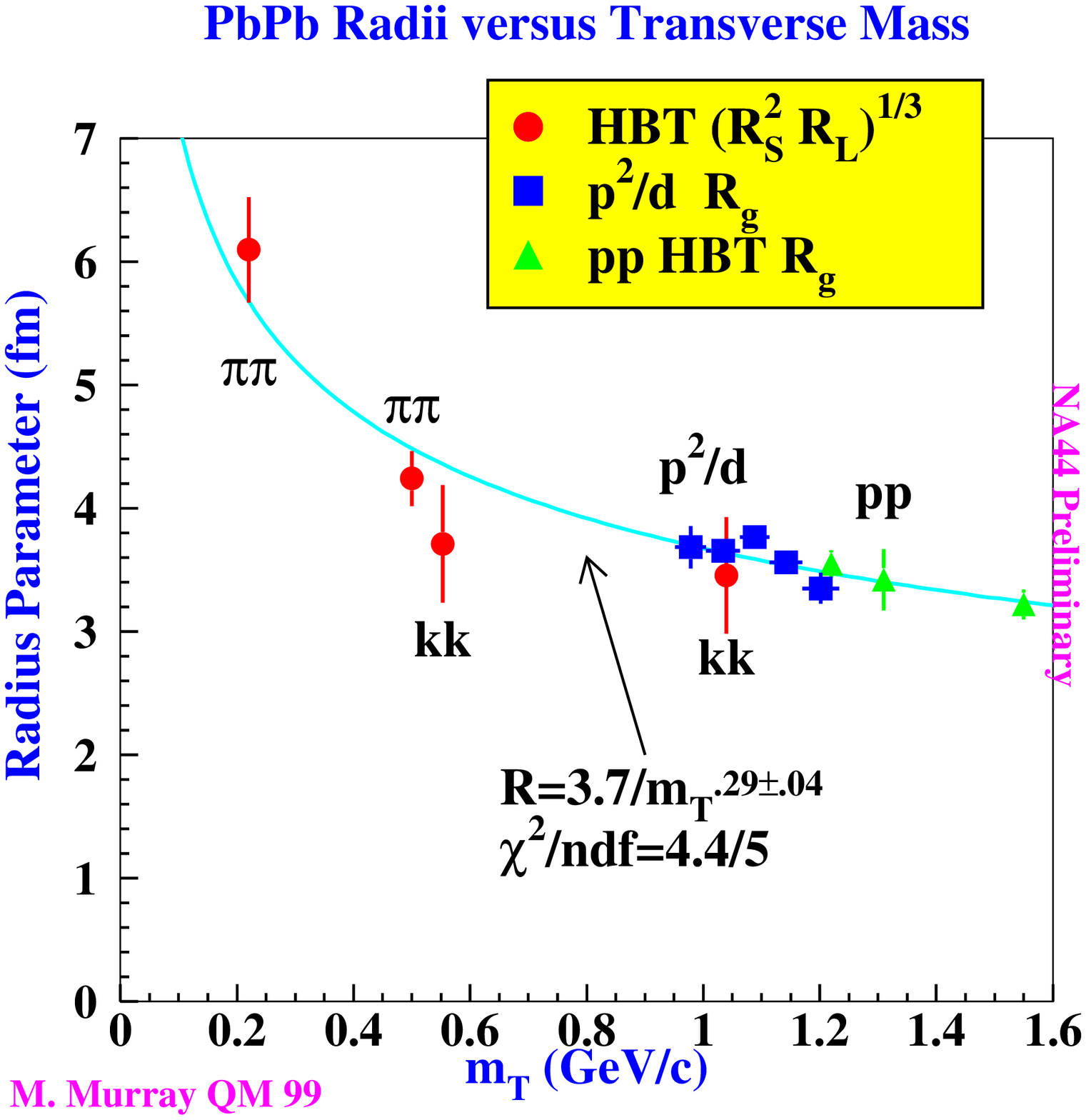}
\end{minipage}
\hfill\hspace*{-1.5cm}
\parbox[c]{6.7cm}{\vspace*{-0.5cm}
    Figure~4. Effective radius $R_{\rm eff}$ of the emission region 
    extracted from $\pi\pi$ and $KK$ Bose-Einstein (HBT) correlations, 
    from $pp$ final state interaction correlations and from the 
    deuteron coalescence probability $d/p^2$, as a function of the 
    transverse mass $M_T\equiv M_\perp$ of the particle pair. The 
    HBT correlations were analyzed in 3 dimensions, and $R_{\rm eff}$
    was defined by the geometric mean of the two transverse and the 
    longitudinal size parameters, $R_{\rm eff}= (R_\perp^2R_\parallel)^{1/3}
    \equiv (R_s^2 R_L)^{1/3}$. (Figure taken from \cite{NA44}.) 
\label{F4}}
\vspace*{-0.7cm}

While Eq.~(\ref{Rp}) does not permit to separate $T_{\rm f}$ from $\vp$
either, the correlation between the two parameters in (\ref{Rp}) is
exactly opposite to that provided by the spectral slopes in (\ref{blue})
and (\ref{flow}). Combining them in a simultaneous analysis of spectra 
and correlations \cite{NA49HBT,TWH99} (see Figure 5) allows for a rather 
accurate separation of directed collective and random thermal motion,
yielding $T_{\rm f}\approx 100$\,MeV and $\vp\approx 0.55\,c$. The 
corresponding thermal energy density is only about $\epsilon_{\rm f}
\approx 0.05$\,\Gf. Even when the kinetic flow energy is added this still
implies an expansion of the fireball volume by a factor 40 from $\tau_0$
to $\tau_{\rm f}$. HBT measurements show that the transverse area grows 
by a factor 4--5, consistent with the large transverse expansion 
velocity $\vp$ \cite{TWH99}; the remaing factor 8--10 must come from 
longitudinal growth.

\begin{minipage}[c]{10cm}
\vspace*{0.5cm}
\hspace*{-0.6cm}
  \epsfxsize 8.5cm
  \epsfbox{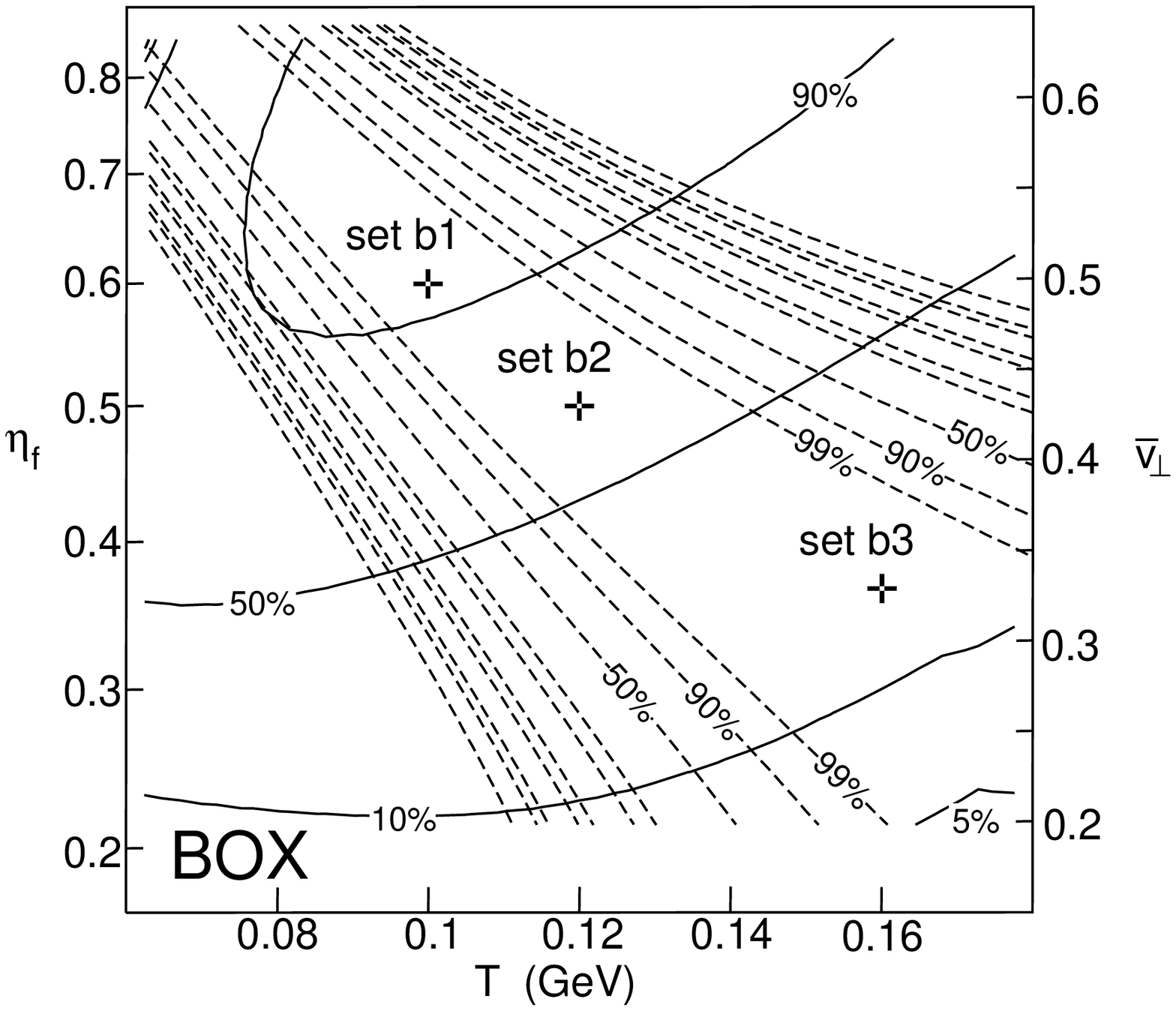}
\end{minipage}
\hfill\hspace*{-1.5cm}
\parbox[c]{6.8cm}{\vspace*{0cm}
    Figure~5. $\chi^2$ contours for the fit of the measured HBT radius 
    parameters from $\pi\pi$ Bose-Einstein correlations \cite{NA49HBT} 
    (widely spaced contours) and of the negative particle spectrum
    \cite{spectra} (narrowly spaced contours) in the rapidity window 
    $3.9 < Y_{\rm lab} < 4.4$ from 158 $A$\,GeV/$c$ Pb+Pb collisions. 
    The best fits require freeze-out temperatures slightly below 
    100\,MeV and average transverse expansion velocities of about 
    0.55\,$c$. For details the reader should refer to \cite{TWH99} 
    from where this figure was taken. 
\label{F5}}
\vspace*{0cm}

\section{THE MISSING RHO: OBSERVING THERMALIZATION AT WORK}

With strong experimental evidence that the Little Bang started out at 
an energy density above 3\,\Gf, but only decoupled at about 
50\,MeV/fm$^3$, we may ask: how can we find out what 
happened in between? Here the $\rho$ meson can provide a first answer:
it can decay into $e^+e^-$ or $\mu^+\mu^-$ pairs which escape from the
fireball without further interactions, and this $\rho$-decay clock ticks
at a rate of 1.3\,fm/$c$, the natural lifetime of the $\rho$. What I 
mean by this is that after one generation of $\rho$'s has decayed, a 
second generation is created by resonant $\pi\pi$ scattering, which can
again decay into dileptons, etc. The number of extra dileptons with the 
invariant mass of the $\rho$ is thus a measure for the time in which 
the fireball consists of strongly interacting hadrons \cite{HL91}. 
Obviously, $\rho$ mesons do not exist before hadrons appear in the 
fireball, so they won't tell us anything about a possible QGP phase 
in its initial stages. But they still allow us to look {\em inside} 
the strongly interacting hadronic fireball at a later stage, still
long before the hadrons decouple.

The experimental check of this conjecture yields a surprise: when the 
CERES/NA45 collaboration looked at the $e^+e^-$ spectrum in 
158\,$A$\,GeV/$c$ Pb+Au collisions (see Figure 6), they could not 
find the $\rho$ at all! Sure, there were extra $e^+e^-$ pairs in the 
mass region of the $\rho$ and below (about 2.5--3 times as many as 
expected), but instead of a nice $\rho$-peak at $m_\rho=770$\,MeV one 
finds only a broad smear \cite{CERES}. Many explanations of the 
CERES-effect have been proposed, but the simplest one consistent 
with the data (for a review see \cite{Rapp}) is {\em collision 
broadening}: there is strong rescattering of the pions, not only 
among each other, but also with the baryons in the hadronic 
resonance gas, and this modifies their spectral densities and, 
as a consequence, leads to a smearing of the $\rho$-resonance in 
the $\pi\pi$ scattering cross section. 

This demonstrates that, after first being formed in the hadronization 
process, the pions (the most abundant species at the SPS) undergo 
intense rescattering before finally freezing out. And this 
again is the mechanism which allows the fireball to reach and maintain 
a state of approximate local thermal equilibrium, to build up 
thermodynamic pressure and to collectively explode, as seen from 
the above analysis of the freeze-out stage. That the dileptons from
collision-broadened $\rho$'s outnumber those from the decay of 
unmodified $\rho$'s emitted at thermal freeze-out (which should show 
up as a normal $\rho$-peak) shows that the hadronic rescattering 
stage must have lasted several $\rho$ lifetimes. 

\begin{minipage}[c]{10cm}
\vspace*{-0.8cm}
\hspace*{-0.6cm}
  \epsfxsize 8.5cm   \epsfysize 9cm 
  \epsfbox{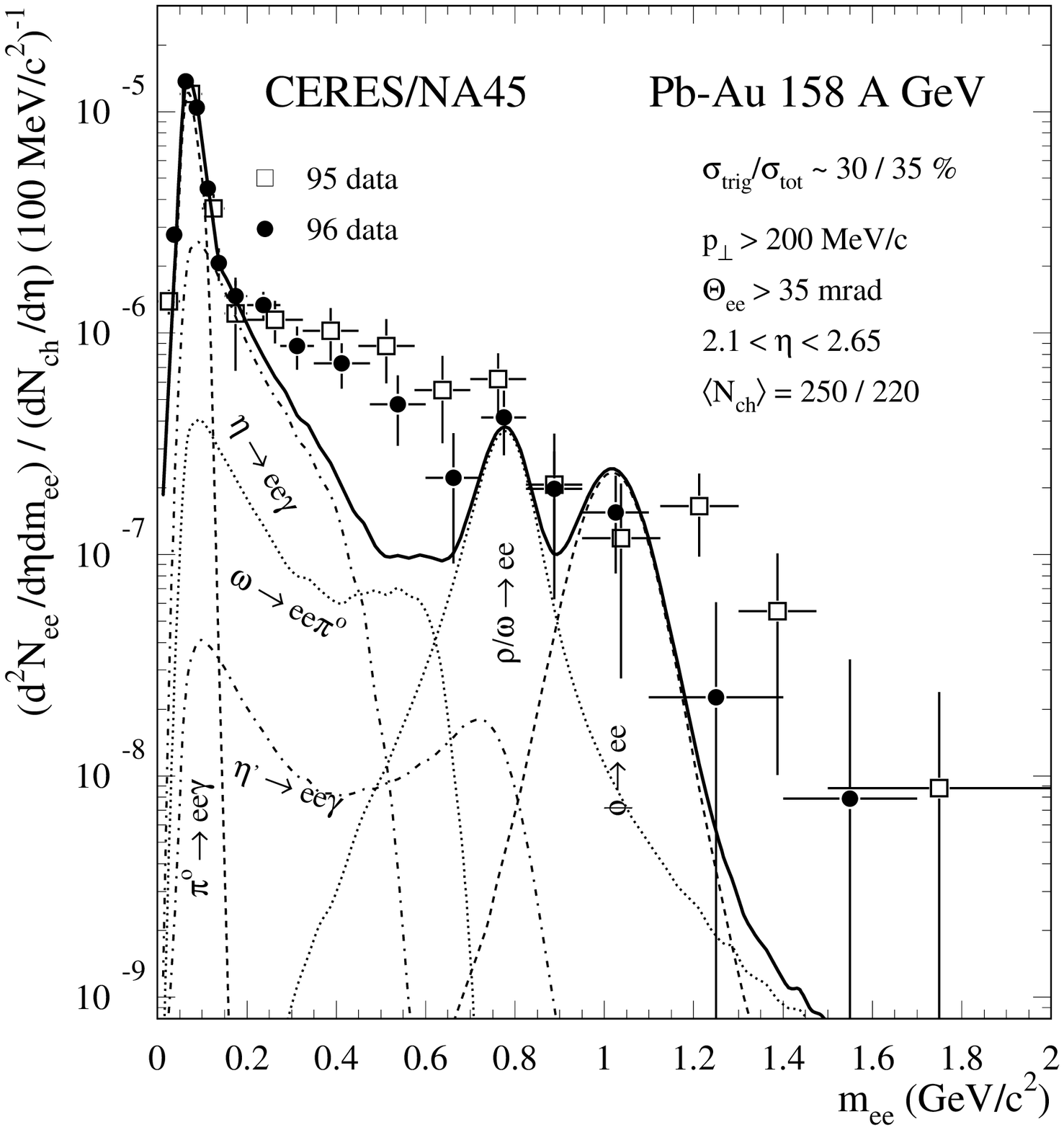}
\end{minipage}
\hfill\hspace*{-2cm}
\parbox[c]{7cm}{\vspace*{0cm}
    Figure~6. Invariant mass spectrum of $e^+e^-$ pairs from 
    158\,$A$\,GeV/$c$ Pb+Au collisions \cite{CERES}. The solid line
    is the expected spectrum (the sum of the many shown contributions)
    from the decays of hadrons produced in $pp$ and $pA$ collisions 
    (where it was experimentally confirmed \cite{CERES}), properly 
    scaled to the Pb+Au case. Two sets of data with different analyses 
    are shown. Note that the $\rho$-peak reappears if only $e^+e^-$ 
    pairs with $p_\perp>500$\,MeV/$c$ are selected \cite{CERES}; such 
    fast $\rho$'s escape quickly from the fireball and are not as 
    strongly affected by collision broadening.
\label{F6}}
\vspace*{0.15cm}

\section{SEEING THE QUARK-HADRON TRANSITION} 

In the rest of this talk I will concentrate on observables which in
heavy-ion collisions were found to be drastically different from 
NN collisions but which we now believe cannot be changed quickly 
enough by hadronic rescattering during the time available between 
hadronization and kinetic freeze-out. Observables for which this 
last property can be firmly established yield insights about where
heavy-ion collisions differ from NN collisions already {\em before 
or during} hadronization, irrespective whether or not the hadrons 
rescatter with each other after being formed. 

Of course, the formation of a quark-gluon plasma is one possibility
how the early stage of a heavy-ion collision may differ from that in 
a NN collision. It is thus important to review a few key QGP 
predictions and check how they fare in comparison with the data. In the 
present Section I discuss {\em strangeness enhancement} as a QGP 
signature, returning to two further QGP predictions in the following 
two Sections. 

{\em Strangeness enhancement and chemical equilibration} was one of
the earliest predicted QGP signatures \cite{Raf}. The idea is simple:
color deconfinement leads to a large gluon density which can create
$s\bar s$ pairs by gluon fusion, and chiral symmetry restoration
makes the strange quarks relatively light, thus reducing the production 
threshold (not to mention that in the QGP strange quarks can be created 
without the need for additional light quarks to make a hadron). The two
effects together should cause a significant reduction of the time scale 
for strangeness saturation and chemical equilibration, compared to 
hadronic rescattering processes after hadronization where the production 
of strange hadron pairs with opposite strangeness is suppressed by large 
thresholds and small cross sections. Since the production of strange 
hadrons in NN and $e^+e^-$ collisions is known to be significantly 
suppressed relative to the expectation from simple statistical 
phase-space considerations \cite{Becattini}, this should lead to a 
relative {\rm enhancement} of strangeness production in heavy-ion 
collisions.

Kinetic simulations, based on known hadronic properties and 
interaction cross sections, have now convincingly shown that it is
not possible to create a state of hadronic chemical equilibrium 
and a significant amount of strangeness enhancement out of a 
non-equilibrium initial state by purely hadronic rescattering (for 
a recent review see \cite{Hqm99}). If you want to get those features
out, you have to put them in at the beginning of the simulation. 

There may be many different ways of doing so. However, the
most efficient way of creating a state of (relative or absolute) 
hadronic chemical equilibrium appears to be provided by the 
hadronization process itself: due to color-confinement, the coalescence 
of colored quarks into hadrons is a process with very large cross 
sections, allowing for many different arrangements among the quarks 
with essentially equal probability. If before hadronization the quarks
and gluons are essentially uncorrelated (like in a QGP), then the most 
likely outcome of the hadronization process is a statistical occupation
of the hadronic phase-space, i.e. a state of maximum entropy, subject 
only to the constraints of conservation of energy, baryon number, net 
strangeness, and {\em (sic!)} the total number of available $s\bar s$ 
pairs. Thus, if (as predicted for the QGP \cite{Raf}) the number of 
$s\bar s$-pairs is enhanced {\em before} the onset of hadronization, or 
by the fragmentation of gluons {\em during} hadronization, their 
statistical distribution over the available hadronic channels will 
naturally lead to an apparent hadronic chemical equilibrium state 
(with the appropriate enhancement of, say, the $\bar \Omega$) 
{\em even if none of the hadrons ever scattered with each other after 
being formed}.

\begin{minipage}[c]{10cm}
\vspace*{-0.6cm}
\hspace*{-0.6cm}
  \epsfxsize 10cm \epsfysize 8cm
  \epsfbox{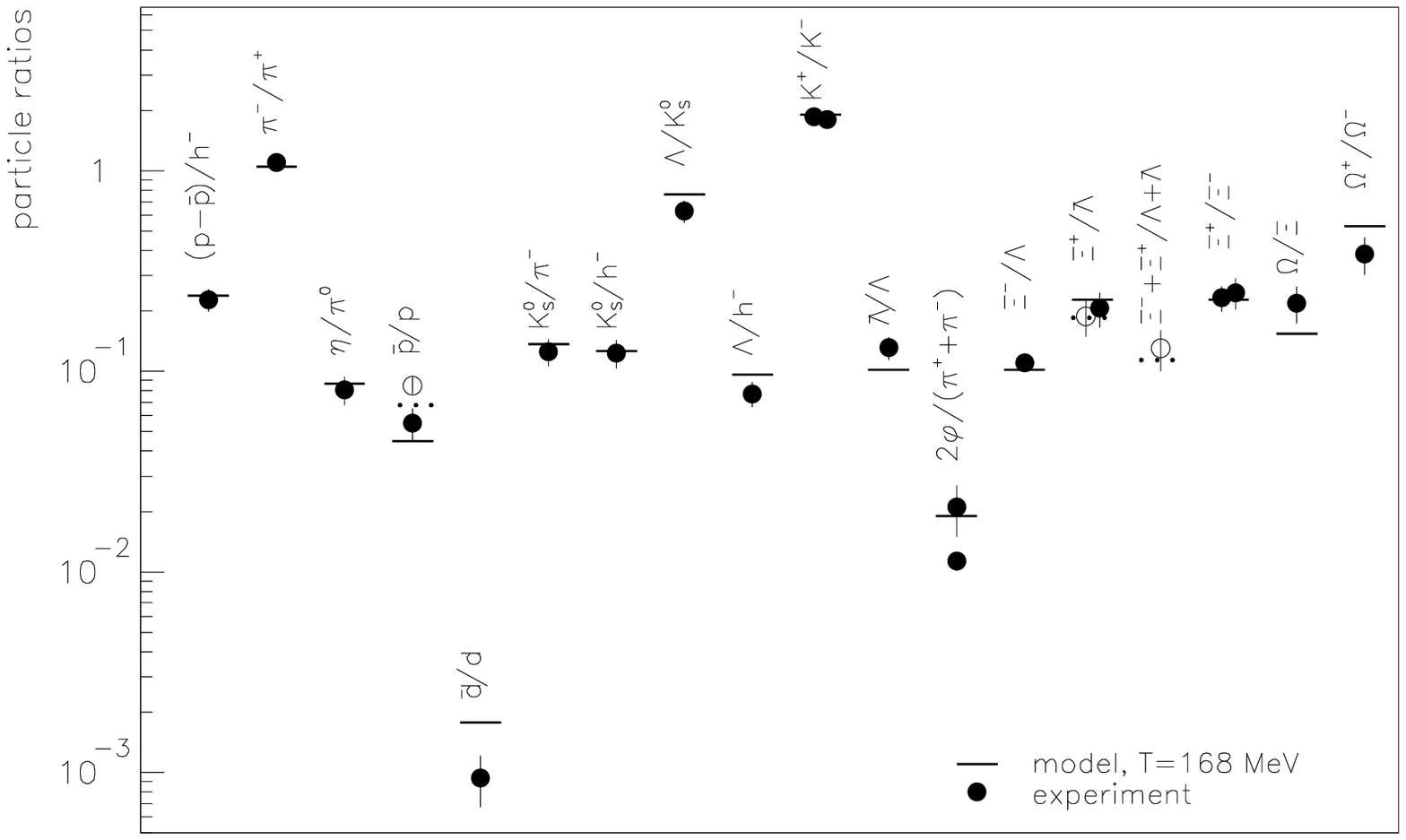}
\end{minipage}
\hfill\hspace*{-1.5cm}
\parbox[c]{6.5cm}{\vspace*{0cm}
    Figure~7. A compilation of measured particle ratios from 
    158\,$A$\,GeV/$c$ Pb+Pb collisions, compared with a hadron resonance
    gas in complete chemical equilibrium (full strangeness saturation)
    at $T_{\rm chem}=168$\,MeV and $\mu_{\rm B}=266$\,MeV \cite{PBM}. 
\label{F7}}
\vspace*{-0.7cm}

Such a state of ``apparent'' or ``pre-established'' chemical equilibrium 
is indeed seen in the experiments: Figure\,7 shows some 18 hadronic 
particle ratios measured in 158\,$A$\,GeV/$c$ Pb+Pb collisions, 
compared with a chemical equilibrium fireball model at 
$T_{\rm chem}=168$\,MeV and $\mu_{\rm B}=266$\,MeV \cite{PBM}. The 
agreement between the data is at least as good as between different 
experiments. The value of $T_{\rm chem}$ is interesting:
in the maximum entropy sense it characterizes the energy density at 
which hadronization occurs (about 0.5\,\Gf) and coincides within errors 
with the critical temperature for color deconfinement from lattice QCD.
If the hadrons were formed by hadronization of a prehadronic state at 
the critical energy density $\epsilon_{\rm c}$ and their abundances froze
out at $T_{\rm chem}=168$\,MeV, there was indeed no time to achieve this
equilibrium configuration by hadronic rescattering; the hadrons must have
been ``born'' into chemical equilibrium \cite{Hqm97,Stock}.

\begin{minipage}[c]{10cm}
\vspace*{-0.8cm}
\hspace*{0.25cm}
  \epsfxsize 9cm \epsfysize 8.5cm
  \epsfbox{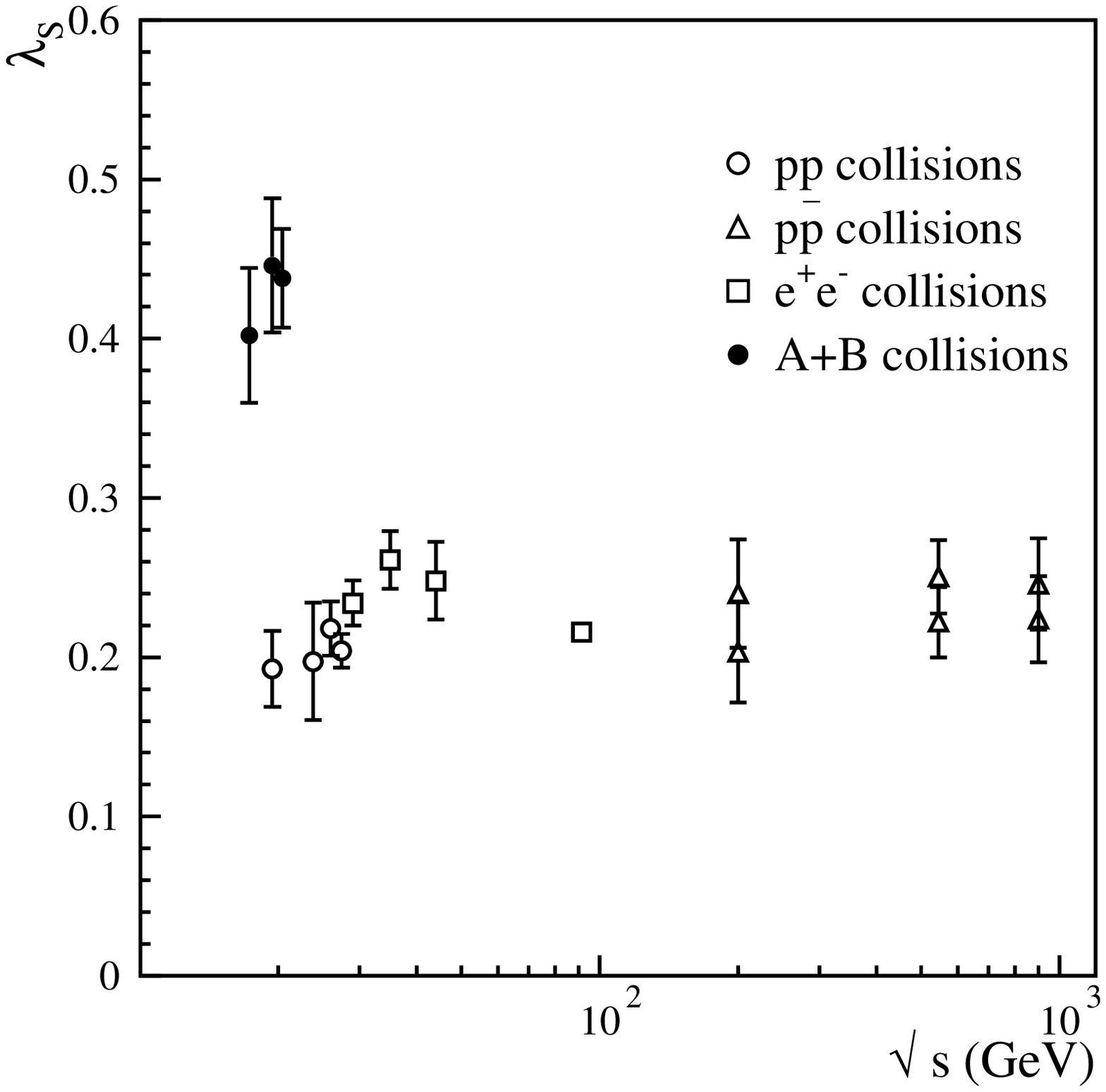}
\end{minipage}
\hfill\hspace*{-1.5cm}
\parbox[c]{6.5cm}{\vspace*{-1.5cm}
    Figure~8. The strangeness suppression factor of produced strange 
    vs. nonstrange valence quarks, $\lambda_s
    = 2\langle \bar s s\rangle/\langle \bar u u + \bar d d\rangle
    \vert_{\rm produced}$, in elementary particle and heavy-ion 
    collisions as a function of $\sqrt{s}$ \cite{BGS}. The two points 
    each for $p\bar p$ collisions reflect the inclusion (exclusion) 
    of the initial valence quarks.
\label{F8}}
\vspace*{-1.2cm}

But this only half the story. Namely, a similar picture of statistical 
hadronization at the critical energy density $\epsilon_{\rm c}$ arises 
even from an analysis of $e^+e^-$, $pp$ and $p\bar p$ collisions 
\cite{Becattini}. What is really dramatically different in heavy-ion 
collisions is the level of strangeness saturation reflected in the
apparent chemical equilibrium state: Figure\,8 shows that the overall 
fraction of strange particles is about twice as high in heavy-ion 
collisions as in elementary particle collisions! Essentially the 
strangeness suppression observed in NN collisions has disappeared in
AA collisions. According to the preceding paragraph, this extra 
strangeness cannot have been produced by final state rescattering; 
it thus reflects the properties of the prehadronic state before 
hadronization. This points to a new, fast strangeness production 
mechanism before hadronization, as predicted for QGP \cite{Raf}.

\begin{minipage}[c]{10cm}
\vspace*{-0.1cm}
\hspace*{-0.5cm}
  \epsfxsize 9cm \epsfysize 5.9cm
  \epsfbox{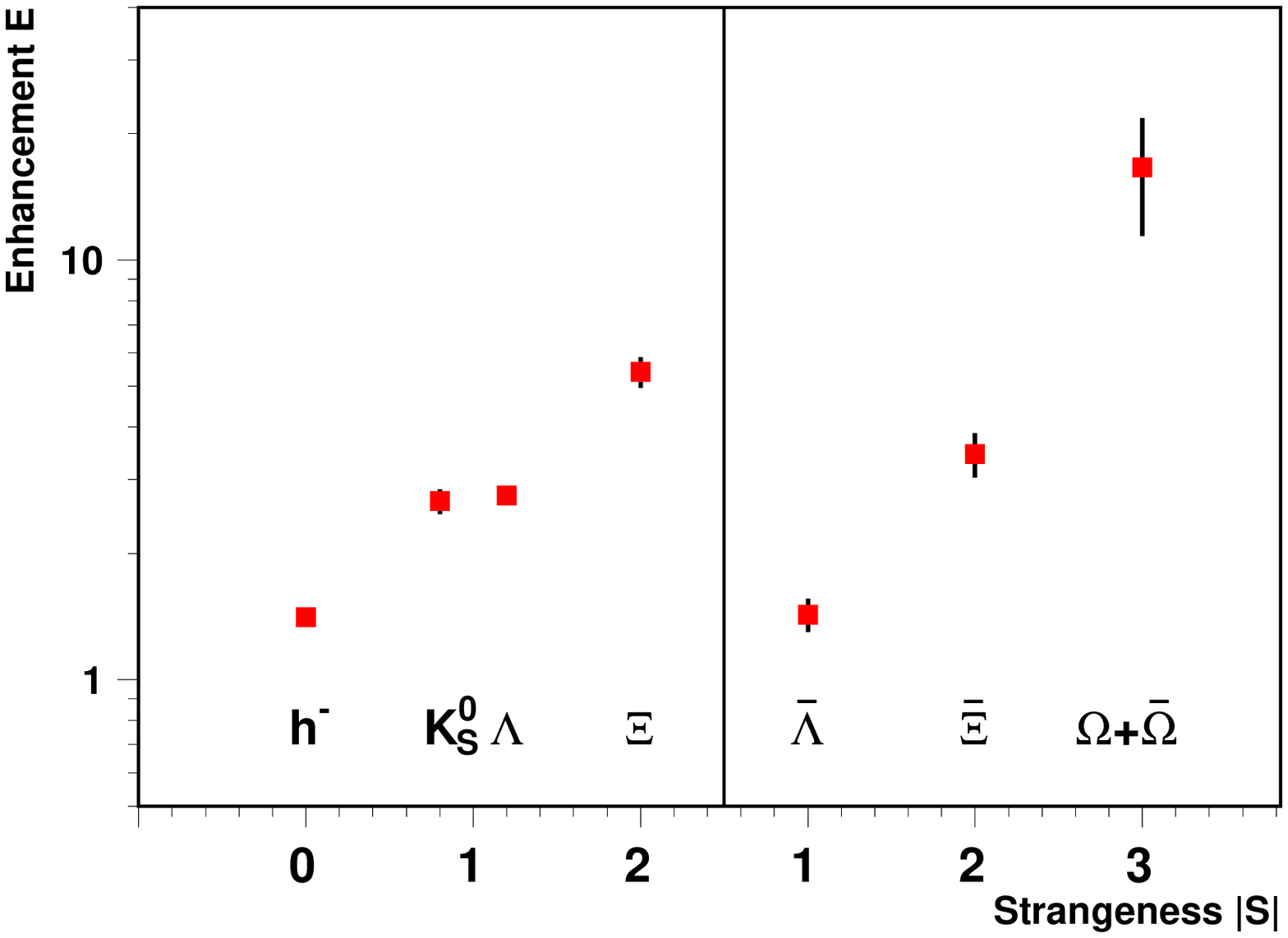}
\end{minipage}
\hfill\hspace*{-1.5cm}
\parbox[c]{6.5cm}{\vspace*{-0.5cm}
    Figure~9. Enhancement factor for the midrapidity yields per 
    participating nucleon in 158\,$A$\,GeV/$c$ Pb+Pb relative to
    p+Pb collisions for various strange and non-strange hadron 
    species \cite{Lietava}. 
\label{F9}}
\vspace*{0.1cm}

A very striking way of plotting these findings is shown in Figure\,9
\cite{Lietava}: relative to $p$+Pb collisions, the number of produced
strange hadrons per participating nucleon is the more strongly enhanced 
the more strange (anti)quarks for its formation are required. For $\Omega$
and $\bar\Omega$ this enhancement factor is about 15! The tendency 
shown in Figure\,9 is completely counterintuitive for hadronic 
rescattering mechanisms, where multistrange (anti)baryons are 
suppressed by higher thresholds than kaons and $\Lambda$'s; but 
it is perfectly consistent with a statistical hadronization picture 
\cite{Bialas} where multi-strange particles profit more from the 
global strangeness enhancement than singly strange hadrons.  

Kinetic codes like URQMD \cite{Bass} tend to lose (anti)baryons (and 
thus a fraction of the initial enhancement of multi-strange 
(anti)baryons) by baryon-antibaryon annihilation during the 
rescattering stage. It was recently shown \cite{RS00} that this is to
a large extent a manifestation of the lack of detailed balance in the 
codes which include processes like $\bar p p\to n\pi$ (with $n=5-6$)
but not their inverse. Rapp and Shuryak \cite{RS00} argue that, as the 
system cools below $T_{\rm chem}$, pions and kaons don't annihilate but 
instead build up a positive chemical potential which enhances the 
probability for the inverse reaction and strongly reduces the net 
annihilation of antibaryons. This is really fortunate, because it
is this lack of abundance-changing processes during the hadronic 
expansion stage which allows us to glimpse the hadronization process
itself through the final hadronic abundances, in spite of intense, 
resonance-mediated {\em elastic} rescattering among the hadrons 
between hadronization at $T_{\rm chem}\approx 170$\,MeV and kinetic 
freeze-out at $T_{\rm f}\approx 100$\,MeV.

\section{$J/\psi$ SUPPRESSION AND COLOR DECONFINEMENT}

The observation of hadronic chemical equilibrium abundances has taken us
from kinetic freeze-out all the way back to the hadronization transition.
The observed strangeness {\em enhancement} gives indirect information 
about the state that existed {\em before} hadronization. It is consistent 
with the hypothesis that this state consisted of color-deconfined 
quark-gluon matter and that intense rescattering among the quarks and 
gluons, before or during the hadronization process, produced the extra
strangeness. But this may not be the only explanation. Can one find other,
perhaps more direct indications for matter containing deconfined gluons 
in the early collision stages? 

This brings us to the second key prediction for QGP formation: Matsui
and Satz suggested \cite{MS86} that the high gluon density resulting
from color deconfinement should Debye-screen the color interaction 
potential between a $c$ and a $\bar c$ quark pair produced during the
initial impact of the two nuclei and thus prevent them from binding
into charmonium states ($J/\psi$, $\chi_c$, $\psi$'). Instead, they 
would eventually find light quark partners to make hadrons with open 
charm. This would lead to a suppression of charmonium production in
heavy-ion collisions, and the screening mechanism should lead to a 
specific suppression pattern which, as a function of energy density 
achieved, first affects the loosely bound $\psi$' and $\chi_c$ states
and then the strongly bound $J/\psi$ ground state \cite{satz}.

There is an expected background to this charmonium suppression which
already exists in $pA$ collisions and can be studied there in order 
to extrapolate it to $AA$: if you put yourself in the CM frame, a 
$c\bar c$ pair created in a hard $pp$ collision (i.e. after $1/(2m_c)
 < 0.1$\,fm/$c$) will be affected by 
interactions with other nucleons from the rest of the nucleus which 
is still sweeping over it. This ``normal suppression'' has by now been
well studied \cite{satz} and is indicated by the straight lines in 
Figure 10. It follows an exponential attenuation with the length $L$ 
of cold nuclear matter of density $\rho_0$ sweeping over the 
$c\bar c$-pair, with an absorption cross section of about 6\,mb for 
{\em both the weakly bound $\psi$' and the tightly bound $J/\psi$}! 
The equal absorption is understood as an effect on the pre-resonant 
$c\bar c$ state at early times, before the bound charmonium states 
actually form (which at SPS energies and above happens only after the
whole nucleus has passed over the pair).

Figure\,10 shows that a deviation from this ``normal'' absorption 
occurs in heavy-ion collisions once the nuclear overlap volume (related
to the variable $L$) becomes sufficiently large. The weakly bound
$\psi$' suffers ``anomalous absorption'' first, at around $L=5$\, fm,
while for the $J/\psi$ and/or the $\chi_c$ (which in $pA$ collisions
is known to contribute about 30\% to the measured $J/\psi$ yield via
its radiative decay $\chi_c\to\gamma\,J/\psi$) the anomalous absorption
does not set in until about $L=7.5$\,fm. 

\vspace*{-0.2cm}
\begin{center}
   \begin{minipage}[t]{7.1truecm}
         \epsfxsize 7.1truecm \epsfbox{f10b.eps}
         \hfill
   \end{minipage}
\hspace{0.5cm}
   \begin{minipage}[t]{7.1truecm}
         \epsfxsize 7.1truecm \epsfysize 6.6cm \epsfbox{f10a.eps}
         \hfill
   \end{minipage}
\end{center}
\vspace*{-0.5cm}
  Figure\,10. NA38/NA50 data on $\psi'/DY$ and $(J/\psi)/DY$ 
  \protect\cite{NA50}, plotted as a function of the nuclear 
  thickness parameter $L$ (see text). 
\label{F10}

\vspace*{-0.3cm}
\hspace*{-1.2cm}
\begin{center}
   \begin{minipage}[t]{7.5truecm}
         \epsfxsize 7.5truecm \epsfysize 7truecm \epsfbox{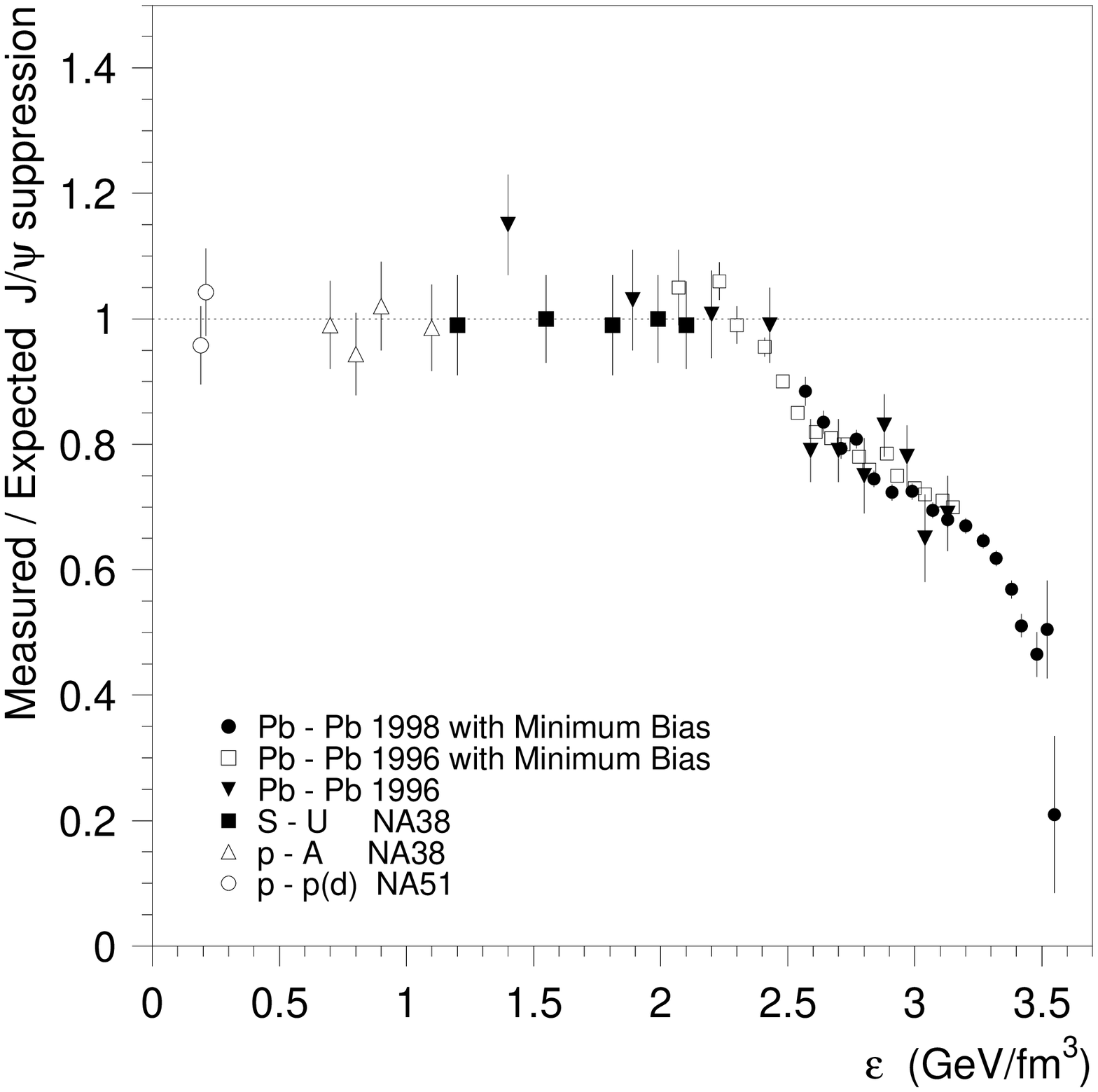}
         \hfill
   \end{minipage}
   \begin{minipage}[t]{7.5truecm}
         \epsfxsize 7.5truecm \epsfysize 7truecm \epsfbox{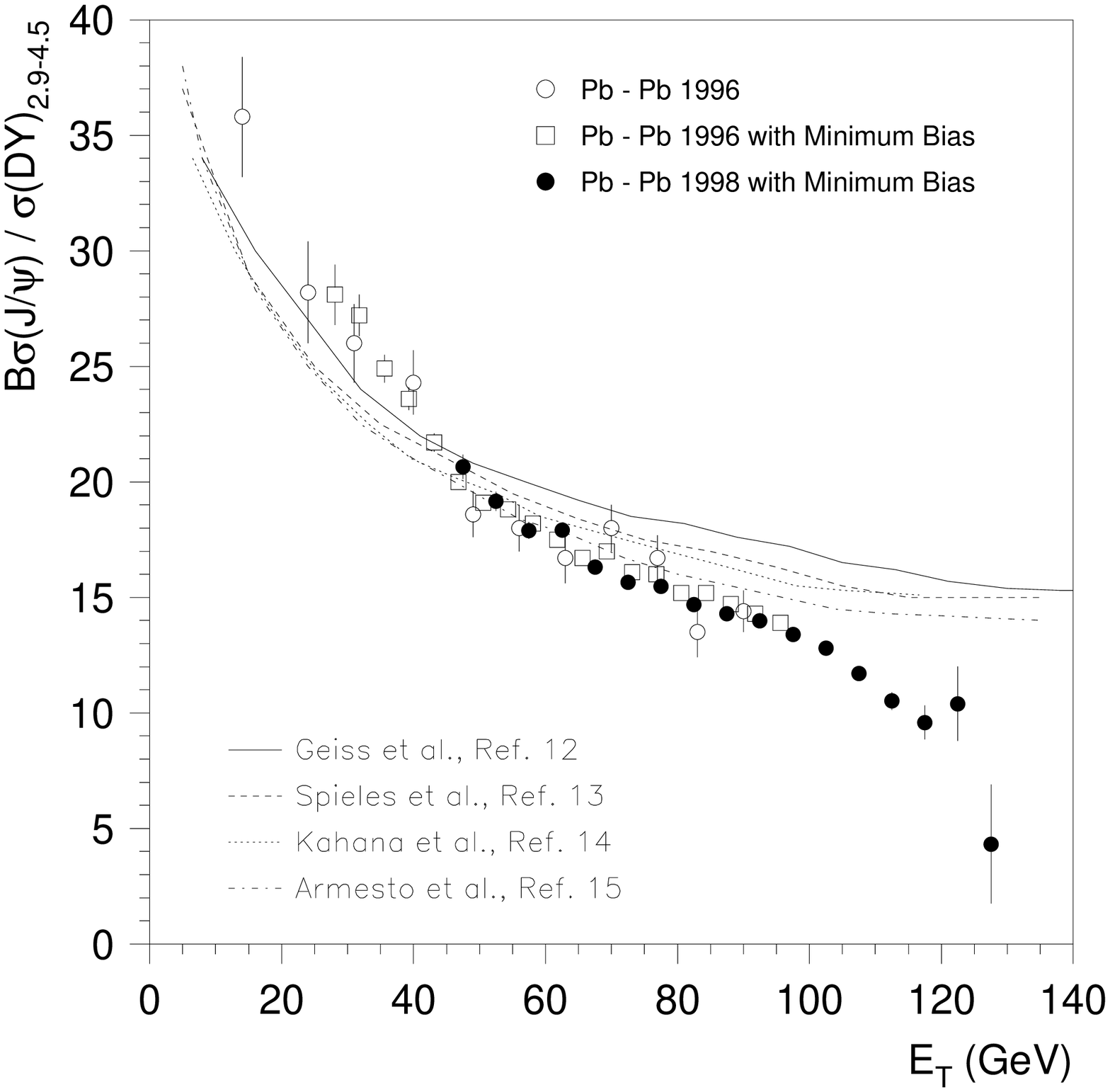}
         \hfill
   \end{minipage}
\end{center}
\vspace*{-0.7cm}
  Figure\,11. Left: ``Anomalous $J/\psi$ suppression'' (see text)
  as a function of initial energy density \cite{NA50eps}. Right: The 
  ratio $(J/\psi)$/Drell-Yan production, as a function of the measured 
  transverse energy $E_T$, compared to different hadronic comover 
  models \cite{NA50eps}.
\label{F11}
\vspace*{0.2cm}

The $J/\psi$ suppression has in the meantime been studied in much 
greater detail as shown in Figure\,11. The left part shows the
``anomalous suppression'' (with the normal pre-resonant absorption 
on the incoming nuclei divided out) as a function of energy density
\cite{NA50eps}. The latter is computed from the measured transverse 
energy $E_T$ via Bjorken's formula (\ref{bj}), where the effective
overlap area in the denominator is related to $E_T$ via a geometric
model and Glauber theory \cite{NA50eps}. The observed suppression 
pattern is interesting: it occurs in two ``waves'', with an 
intermediate flattening after about 30\% of suppression (which one
might be tempted to associate with the complete suppression of
the $\chi_c$ component) followed by a stronger suppression at energy 
densities above 3\,\Gf. 

The real origin of this ``wavy'' structure is not yet clear; 
on a superficial level it is qualitatively consistent with
a suppression hierarchy $\psi'\to\chi_c\to J/\psi$ as expected 
from the QGP color-screening scenario. In any case, the right part 
of Figure\,11 shows that it can not be reproduced
by conventional hadronic final state interactions (collision
dissociation) between the charmonium and the produced hadrons.
Such ``hadronic comover models'' must be tuned to relatively large
inelastic $J/\psi$-comover scattering cross sections to even reproduce
the average suppression in Pb+Pb collisions, at the expense of sometimes
overpredicting the effect in S+U collisions. Even then most of the
destructive interactions happen at very early times where the comover
densities are so high that a hadronic description should really not 
yet apply. Most importantly, however, they all consistently fail 
to reproduce the strong suppression at very high $E_T$.

It was recently noted that the apparent onset of the second, stronger
suppression above $\epsilon=3$\,\Gf coincides roughly with the ``knee'' 
in the $E_T$-distribution, i.e. with the point of full nuclear overlap 
above which one enters the region of $E_T$ fluctuations \cite{Cap,Bl}. 
These fluctuations tend to increase the $J/\psi$ suppression even in 
the comover models, but the effect is not strong enough to reproduce 
the data \cite{Cap}. On the other hand, if 
one associates them with fluctuations of the energy density in the QGP 
Debye-screening approach the fluctuation effects are much stronger and the 
data can be reproduced \cite{Bl}. It would be important to clarify this 
issue, e.g. by performing U+U collisions where, due to the deformation 
of the uranium nucleus, in the tip-on-tip configuration similar energy 
densities can be reached without having to exploit $E_T$ fluctuations.

\section{THERMAL ELECTROMAGNETIC RADIATION}

The third (and earliest) key prediction for QGP formation was thermal
radiation of (real and virtual) photons from the thermalized quarks in 
the QGP \cite{Sh}. They are conceptually clean direct probes of the QGP 
phase itself, but experimentally difficult due to large backgrounds. 
Nonetheless, experiments at the SPS have searched for this type of 
radiation. An excess yield in the photon spectrum above about 1.5 GeV
which is attributed to direct emission was found by WA98 \cite{WA98}
and discussed by Th. Peitzmann at this meeting. It is, I think, fair to
say that the relation of this signal with thermal QGP radiation is
presently unclear. The NA50 Collaboration found an excess in the 
dimuon spectrum between the $\phi$ and $J/\psi$ peaks which was 
interpreted by them as an excess in open charm production \cite{NA50exc} 
but which may also be due to thermal QGP radiation \cite{K}. Again,
the true origin is not clear. The situation is such that, for the 
initial temperatures reachable at the SPS, the theoretical predictions
for a thermal radiation signal are so low that it is not obvious that
it can be dug out from the experimental background. Since the thermal 
radiation rate goes with $T^4$, the higher initial temperatures 
reachable at RHIC and LHC, combined with the longer lifetime of the 
plasma phase, should help a lot to see the plasma ``shine'' and 
thus confirm this prediction experimentally.

\section{CONCLUSIONS}

Relativistic heavy-ion collisions at the CERN SPS have taken us into new 
and unprecedented regions of energy density: in Pb+Pb collisions
at 158\,$A$\,GeV/$c$ initial energy densities of about 3.5\,\Gf 
(i.e. more than 20 times the energy density of cold nuclear matter)
have been created over large volumes. If the matter was approximately
thermalized even at this early stage (for which we do not yet have
convincing direct evidence, although data on elliptic flow 
\cite{ellip,kolb}, which I had no time and space to discuss, point 
in this direction), the initial temperature was around 210--220 MeV, 
i.e. 30\% above the critical temperature for color deconfinement.

We have strong and direct experimental evidence for a large degree of
thermalization and strong collective behaviour in the late stages
of the collision, driven by intense rescattering among the fireball
constituents which is directly visible in the low-mass dilepton spectra.
At kinetic freeze-out the fireball radiates hadrons with a temperature 
of about 100 MeV, at the same time undergoing collective explosion with 
more than half the light velocity (the ``Little Bang'').

Extensive theoretical simulations have shown that conventional hadronic
processes during the hadronic rescattering phase lead mostly to elastic 
collisions and are very inefficient in changing the final hadron 
abundances. The observed hadronic particle ratios thus reflect the 
``primordial hadrosynthesis'' in the Little Bang and provide a direct 
glimpse of the hadron formation stage. The data show that the hadrons 
are born into a state of ``pre-established chemical equilibrium'' at a
temperature of about 170\,MeV which coincides with the deconfinement 
temperature predicted
by lattice QCD. This is the first observation of thermal equilibrium 
matter at such a high temperature and energy density. The phenomenon is
most easily understood in terms of statistical hadronzation of a QGP,
although other mechanisms with similar statistical features cannot be 
excluded. The experimentally determined ``chemical'' and ``thermal''
freeze-out points can be connected by an isentropic expansion trajectory
as indicated in Figure\,1. (Note that the part of the same trajectory
which runs through the QGP phase is so far speculative.)

Strangeness enhancement and charmonium suppression have been predicted
as QGP signatures and are indeed found to characterize the prehadronic 
state from which the observed hadrons appear during hadronization. These
features cannot be understood in terms of conventional hadronic final 
state rescattering effects after hadronization. They are  
consistent with QGP expectations, although other explanations may 
still be possible. The inability to understand these properties
in terms of known hadronic physics warrants the characterization of
this prehadronic state as a ``new state of matter'' \cite{CERN}.

{\em So what is missing to claim ``discovery'' of the quark-gluon plasma?} 
First, on the theoretical side, we only know that with known hadronic 
physics we {\em can not} describe the data, but it has not yet been 
shown that a fully dynamical theory which begins with QGP and follows 
the system until freeze-out actually {\em can} describe all observations. 
One reason is that a description of strongly interacting matter and its 
dynamics in the neighborhood of the phase transition is an 
exceedingly difficult problem, and that at the SPS we are never far 
away from the phase transition. This task may thus become easier at 
RHIC/LHC than at the SPS. Furthermore, the evolution of the 
``late signatures'' from $pp$ via $pA$ to $AA$ collisions, to establish 
a clear line between ``conventional hadronic physics'' and 
``new physics'' has so far not received enough careful theoretical 
attention. This should be remedied, but it requires much improved 
$pp$ and $pA$ data at the same energy, with more differential 
experimental information. Such data can (only) be obtained at the SPS.

Further important experimental questions which can be answered at the 
SPS (and in a few cases only there) are: Assuming that we have seen 
quark deconfinement, where is its energy threshold? How big does the 
collision system have to be to establish approximate thermal equilibrium 
and strangeness saturation and to exhibit collective flow? If charmonia 
are suppressed, what happens with open charm -- is Drell-Yan production 
of dileptons (which depends on the quark structure functions in the 
colliding nuclei) really the appropriate normalization for $J/\psi$ 
suppression (which is sensitive to the gluon structure functions 
since $c \bar c$ pairs are made from gluons)? Some answers will be 
provided by data already collected at lower beam energies and with 
smaller nuclei and more peripheral collisions. A search for open 
charm at the SPS was proposed \cite{NA60} and recently approved.
But other questions are still waiting for proposals to be made.

Still, with the SPS we will never be able to get ``away from the edge''. 
A detailed characterization of the ``new state of matter'' will only be 
possible when the larger initial energy densities and resulting longer 
plasma lifetimes before hadronization provided by RHIC and LHC become 
available. The higher initial temperatures are expected to give an 
observable thermal radiation signal, thus allowing to measure the 
thermal ``QGP structure functions''. The higher collision energies 
allow for the creation of high-$p_t$ jets which can then be used as 
probes of the QGP, by their interactions with the plasma when they 
penetrate it. And last not least, the fact that RHIC is a dedicated 
heavy-ion machine at which experiments can be run for a large 
fraction of each year is conducive to the systematic studies that 
will be required to eventually obtain a complete and consistent
picture of quark-gluon plasma dynamics.


\end{document}